\documentclass[aps,pra,twocolumn,showpacs,floatfix]{revtex4}

\usepackage{amssymb,amsmath}
\usepackage{graphicx}
\def\ud{\mathrm{d}}

\def\oP{\omega_{P}}
\def\oC{\omega_{C}}
\def\oR{\omega_{R}}
\def\DA{\Delta_{A}}
\def\DC{\Delta_{C}}

\def\nq{_{n,q}}
\def\logneg{E_{\mathcal{N}}}

\newcommand{\mat}[1]{\mathbf{#1}}

\begin{document}
\title{Quantum noise of a Bose-Einstein condensate in an optical cavity, correlations and entanglement}
\author{G.\ Szirmai$^{1,2}$}
\author{D.\ Nagy$^{1}$}
\author{P.\ Domokos$^{1}$}
\affiliation{$^{1}$Research Institute for Solid State Physics and Optics, Hungarian Academy of Sciences, P.O. Box 49, H-1525 Budapest, Hungary}
\affiliation{$^{2}$ICFO-Institut de Ci\`encies Fot\`oniques, 08860 Castelldefels (Barcelona), Spain}

\begin{abstract}
A Bose-Einstein condensate of ultracold atoms inside the field of a laser-driven optical cavity exhibits dispersive optical bistability. We describe this system by using mean-field approximation and by analyzing the correlation functions of the linearized quantum fluctuations around the mean-field solution.  The entanglement and the statistics of the atom-field quadratures are given in the stationary state. It is shown that the mean-field solution, i.e. the Bose-Einstein condensate is robust against entanglement generation for most part of the phase diagram.
\end{abstract}


\maketitle

\section{Introduction}
\label{sec:intro}

Nonlinear open systems often produce bistabilities or dynamical phase transitions. A nice example is the behavior of a Bose-Einstein condensate (BEC) in a pumped high-finesse optical cavity, where the nonlinearity is produced by the dispersive atom-light interaction \cite{brennecke07,colombe07,brennecke08a,ritter09a,murch08a,gupta07a,klinner06,slama:063620}. The weak cavity pumping causes a classical electromagnetic field to build up between the mirrors. The atoms coupled dispersively to the radiation field detune the cavity according to the overlap of their spatial distribution with the mode function of the electric field. Consequently, the cavity resonance frequency can be shifted away from or towards  the frequency of the pumping laser; thereby a big variation in intensity can be induced merely by the spatial redistribution of the atoms. In turn, the intensity change translates into the variation of the depth of the optical dipole potential, and so it acts back upon the atomic distribution itself. In a tiny region of the parameter settings close to the cavity resonance two stable or metastable configurations can exist, giving rise to a dynamical phase transition.

Atom-light interaction itself is a major issue in the studies of ultracold quantum gases, being the most universal tool in accessing the properties of the system either by slowing the cloud of atoms, trapping them in classical potentials, putting obstacles in their path, or finally detecting and imaging them. Moreover, recent proposals have raised the possibility of quantum state preparation of the atomic ensemble with the help of measuring the output photon signal of a pumped optical resonator \cite{mekhov07a,mekhov09a,mekhov09b,chen:023812,chen:043801}. The cornerstone of such a quantum state preparation with the help of a quantum nondemolition measurement is also the mutual backaction between the atomic and photonic degrees of freedom. The study of correlations between atomic motion and light generated by atom-light interaction in an optical cavity is therefore of fundamental importance.

An important research area on the manifestation of light-matter interaction is optomechanics, where the radiation pressure force of a single mode Fabry-P\'erot resonator is used to manipulate the center of mass motion of a mechanical oscillator. For a short review on optomechanical systems and on their experimental realizations see Ref. \cite{marquardt09}, and also references therein. The reason of popularity of optomechanical studies, besides experimental realizability, is that the theoretical description can be performed relatively simply, involving only the few modes of the cavity field and one mode for the motion of the mirror \cite{law95a,vitali07a,genes08a}. Such a paradigmatic system is an ideal playground to test correlations between light and mesoscopic objects, to understand the underlying physics and to speculate on possible applications in quantum information processing.  

In recent experiments done with ultracold bosons in optical resonators the above concepts unify nicely \cite{brennecke07,colombe07,brennecke08a,ritter09a,murch08a,gupta07a,klinner06,slama:063620}. The cavity radiation field couples to a single collective motional excitation of the Bose condensed atomic sample. Starting the experiment with a pure Bose-Einstein condensate, other motional excitations can be safely disregarded and so a situation analogous to optomechanics can be realized without a movable mirror, but rather with the collective motion of an ensemble of atoms. The difference between the experimental tools (manipulation and detection methods) of traditional optomechanical systems and those with ultracold gases nicely complement each other, while the theoretical description is very similar. 

The aim of the present paper is to discuss correlations generated in a hybrid system of ultracold bosons and the radiation field, especially close to the region of parameter settings where the system shows bistability and where the optomechanical simplification can be harvested. Photons leaking out of the resonator make the cavity field noisy, which infiltrates the dynamics of atomic motion \cite{szirmai09a,murch08a,nagy09}. In turn, quantum fluctuations of the atom field have a back-action on the photon statistics. Correlated fluctuations of the light and matter wave fields appear then, which are strongly enhanced close to the critical regime of bistability. The study of correlations is further motivated by the need for justifying the basic assumption of the generally used mean-field theories \cite{horak00,horak01b,nagy08,szirmai09a,zhang08a,zhang08b,zhang09a}, namely, that the atom-photon cross correlations are negligible and mean values of the light and atomic operators can be decoupled.

The paper is organized as follows. The backbone of the paper is Sec. \ref{sec:model}, where the model and the theoretical description of the system in mean-field level is presented in great detail. Many of the elements of other theoretical models, e.g., the cavity cooling of BEC excitations \cite{horak01b}, spatial self-organization of a BEC in the cavity \cite{nagy08,gopalakrishnan09a}, and the transient collective atomic recoil lasing \cite{moore99,fallani:033612} will be recapitulated here to give a full account of the mean field dynamics of a BEC in a cavity. The aim, partially, is to reach the optomechanical model and discuss the assumptions and approximations to arrive there. Special attention is paid to the effect of nonlinearities: (i) the nonlinearity caused by atom-light interaction, responsible for the creation of a periodic optical potential and also for an effective atom-atom collective interaction, and (ii) the nonlinearity caused by atomic s-wave collisions. In Sec. \ref{sec:res} we present first mean-field results and compare them to  experimental observations, wherever applicable. Furthermore, the auto- and cross-correlations of the quantum fluctuations of the fields will be calculated in the stationary state formed by the balance of cavity loss and vacuum noise driving. Finally, a summary is given in Sec. \ref{sec:disc}.  

\section{Theoretical description of the system}
\label{sec:model}

The system consists of a single mode, high-finesse optical Fabry-P\'erot resonator with a waist much smaller than the cavity length and a sample of dilute, ultracold bosonic atoms prepared to be Bose-Einstein condensed. The condensate is supposed to be cigar shaped along the cavity axis, with a strong transverse confinement. The radiation field inside the cavity is pumped through one of its mirrors by a laser with frequency $\oP$ and wavenumber $k=\oP/c$, with $c$ being the speed of light. The laser frequency is far detuned from the atomic transitions, so the populations of the electronic excited states are negligible. In this limit the atomic internal degrees of freedom are frozen; and the atom-light interaction is purely dispersive. On the other hand, the cavity frequency $\oC$ lies close to the pump frequency $\oP$: the detuning $\DC=\oP-\oC$ is comparable to $\kappa$, this latter being half of the inverse lifetime of a photon inside the cavity.

\subsection{Hamiltonian}
\label{ssec:ham}

In the frame co-rotating with the pumping laser field the Hamiltonian of the system can be approximated as
\begin{subequations}
\begin{equation}
H=H_A+H_C+H_{AC}+H_{CL}+H_{\mathrm{vac}}.
\end{equation}
$H_A$ is the Hamilton operator of the ground state atoms inside the cavity, given by
\begin{equation}
\label{eq:Hatom}
H_A=\!\int \!\Psi^\dagger(x)\Big[\frac{-\hslash^2}{2m}\frac{d^2}{dx^2} + V_{\mathrm{ext}}(x) + \frac{g}{2}\Psi^\dagger(x)\Psi(x)\Big]\Psi(x) dx,
\end{equation}
with $m$ being the mass of the atoms, $V_{\mathrm{ext}}(x)$ is the external confining potential along the cavity axis and $g$ being the s-wave scattering constant in 1-dimension. 
The term $H_C$ of the Hamiltonian describes the radiation field of the empty, single-mode cavity,
\begin{equation} 
\label{Hcav}
H_C = -\hslash\DC\, a^\dagger a.
\end{equation}
The dispersive interaction between the cavity radiation field and the atoms in this low excitation limit is given by the AC-Stark shift, or light shift:
\begin{equation} 
\label{HAC}
H_{AC} = \hslash U_0\, a^\dagger a \int \Psi^\dagger(x)\Psi(x)  \cos^2(k\,x) dx,
\end{equation}
with $U_0$ being the single atom light-shift, $U_0=g_{CA}^2/\DA$; the unique longitudinal mode function of the single mode cavity is $\cos(k\,x)$ with wavenumber $k=\oP/c=2\pi/\lambda$. The part describing the coupling of the cavity field to that of the pump laser is given by
\begin{equation} 
\label{HCL}
H_{CL}=\eta^*\,a + \eta\,a^\dagger,
\end{equation}
\end{subequations}
with $\eta$ being the strength of the driving field; and the asterisk stands for complex conjugation. The last part of the Hamiltonian, $H_{\mathrm{vac}}$ describes the interaction of the cavity field with the broadband reservoir of external radiation field modes via the partially transmissive mirrors. We will give account for this interaction within the Markov approximation, by means of introducing a loss rate $2 \kappa$ and a Gaussian white noise operator $\xi(t)$ in the Heisenberg equation of motion for the field operators \cite{louisell}.

\subsection{Equations of motion}
\label{ssec:eqmo}

The equation of motion of the photonic annihilation operator is given by
\begin{subequations}
\label{eqs:eqmo}
\begin{multline} 
i\frac{\ud}{\ud t}a(t)=\Big[-\DC+\int \Psi^\dagger(x,t)\Psi(x,t)U(x) dx-i\kappa\Big]a(t)\\
 + i \eta + i\xi(t),\label{eq:eqmores}
\end{multline}
with $U(x)=U_0\cos^2(k\,x)$ the local single atom light-shift, which is a periodic function. Its period is noticeably $L=\lambda/2$, since it contains the mode function squared. The operator $\xi(t)$ describes Gaussian white noise with zero mean and with the only non-vanishing correlation
\begin{equation}
\label{eq:noisecorr} 
\left\langle\xi(t)\xi^\dagger(t')\right\rangle=2\kappa\delta(t-t').
\end{equation}
It is nicely exhibited in Eq.\ \eqref{eq:eqmores}, that the dispersive interaction between the atoms and the radiation field causes a shift in the resonator frequency proportional to the atomic density $\Psi^\dagger(x,t)\Psi(x,t)$. However, this frequency shift is an operator and couples the equations of motion of the radiation field to those of the atomic field operators in a nonlinear way. The equation of motion of the atomic field operator reads 
\begin{multline}
i\hslash\frac{\partial}{\partial t}\Psi(x,t)=\Big[-\frac{\hslash^2\Delta}{2m}+V_{\text{ext}}(x)+\hslash a^\dagger(t)a(t)U(x)\\
+ g\Psi^\dagger(x,t)\Psi(x,t)\Big]\Psi(x,t).\label{eq:eqmoat}
\end{multline}
\end{subequations}
In the atomic part of the equation of motion \eqref{eq:eqmoat}, in addition to the inert trap potential $V_{\text{ext}}$, the atom-light interaction creates a $\lambda/2$ periodic optical potential for the atoms, proportional to the dynamical photon number operator $a^\dagger a$.

\subsection{Mean-field solution}

To solve the coupled nonlinear operator equations \eqref{eqs:eqmo} simultaneously is a hard task. The most convenient approximation is the mean-field approximation, when one separates first the operators into a mean value and to fluctuations around it:
\begin{subequations}
\label{eqs:sepmf}
\begin{align} 
\Psi(x,t)&=\sqrt{N}\varphi(x,t)+\delta\Psi(x,t),\\
a(t)&=\alpha(t)+\delta a(t).
\end{align}
\end{subequations} 
The mean values are c-numbers, defined as $\varphi(x,t)=N^{-1/2}\langle\Psi(x,t)\rangle$ the so-called condensate wavefunction, which is normalized to unity; and $\alpha(t)=\langle a(t)\rangle$ the coherent part of the cavity field. Consequently, the fluctuations have zero mean. The time evolution of the mean values is obtained by substituting \eqref{eqs:sepmf} into the equations of motion \eqref{eqs:eqmo} and neglecting all terms containing fluctuations. By this way one arrives at a Gross-Pitaevskii like set of equations of the coupled dynamics
\begin{subequations} 
\label{eqs:gpe}
\begin{align}
i\frac{\ud}{\ud t}\alpha(t)&=\Big[-\DC+N\langle U\rangle - i \kappa\Big]\alpha(t)+i\eta,\label{eq:gperad}\\
i\hslash\frac{\partial}{\partial t}\varphi(x,t)&=\bigg[-\frac{\hslash^2\Delta}{2m}+V_{\text{ext}}(x)+\hslash|\alpha(t)|^2U(x)\nonumber\\
&+g\,N\,|\varphi(x,t)|^2\bigg]\varphi(x,t),\label{eq:gpeat}
\end{align}
\end{subequations}
with the notation $\langle U\rangle\equiv\int\varphi^*(x,t)U(x)\varphi(x,t)dx$.

There is a simplification of the numerical problem due to the possible separation of time scales. The time evolution of Eq.\ \eqref{eq:gperad} is governed by two characteristic frequencies, namely the detuning $\DC$ and the photon loss rate $\kappa$. In Eq.\ \eqref{eq:gpeat} the characteristic frequency is set by the recoil frequency $\oR=\hslash k^2/(2m)$. In experiments the latter one is usually several orders of magnitude smaller than the former ones. For example in the Esslinger group experiments \cite{brennecke07,brennecke08a,ritter09a}, the parameters $|\Delta_C|\sim\kappa\approx2\pi\times1\,\mathrm{MHz}$, while the recoil frequency is about $\oR\approx2\pi\times4\,\mathrm{kHz}$ . In this situation the dynamics of the resonator field relaxes very fast compared to the dynamics of the atomic motion and therefore can be considered instantaneous with respect to the relaxation of the condensate. One can assume for any given atomic configuration that the resonator field has already reached its steady state value, which, by Eq.\ \eqref{eq:gperad}, is
\begin{equation}
\label{eq:adiabel}
\alpha_{\text{ss}}=\frac{i\eta}{\DC-N\langle U\rangle+i\kappa}.
\end{equation}
This steady-state mean field provides the optical potential in Eq.\ \eqref{eq:gpeat}. So in the end one only integrates solely Eq.\ \eqref{eq:gpeat}, with $\alpha(t)$ adiabatically eliminated and inserted from Eq.\ \eqref{eq:adiabel}, instead of the coupled Eqs.\ \eqref{eq:gperad} and \eqref{eq:gpeat}. It is worth noting that if the time scales of the resonator field and BEC dynamics do not differ that much, some complex coupled solutions can exist \cite{marquardt:103901}, which need the simultaneous integration of Eqs.\ \eqref{eq:gperad} and \eqref{eq:gpeat}.

After the adiabatic elimination of the photon field $\alpha$, the most direct method to numerically calculate the steady state of the BEC wavefunction $\varphi(x)$ is the one based on the imaginary-time propagation of Eq.\ \eqref{eq:gpeat}. In real-time the steady-state solution has the time dependence:
\begin{equation}
\varphi(x,t)=\varphi(x)\,e^{-i\mu t/\hslash},
\end{equation}
with $\mu/\hslash$ the lowest frequency eigenvalue of the nonlinear problem \eqref{eq:gpeat}. In imaginary time all fluctuations around the steady-state die out, since they propagate with higher frequencies in real time and consequently vanish faster in imaginary time than the steady-state solution. One just has to renormalize the solution $\varphi(x,t)$ after some finite propagation time. Note that, since all quantities on the left hand side of Eq.\ \eqref{eq:gpeat} are real, the condensate wavefunction $\varphi(x)$ can also be chosen real. We also note at this point that, as we will see later, in the case of an effective blue detuning $\DC-N\langle U\rangle>0$, the resonator field heats the atomic motion (some excitations have positive imaginary parts) and there is no steady-state condensate wavefunction at all. However, due to the method of imaginary time propagation, one can find a BEC wavefunction even in this case, corresponding to a dynamically unstable equilibrium situation.

\subsection{Fluctuations around the mean-field solution}

Having obtained the steady-state values of the BEC wavefunction and the resonator field amplitude, one can look for the fluctuations of the annihilation (and creation) operators $\delta a(t)$, $\delta\Psi(x,t)$ ($\delta a^\dagger(t)$, $\delta\Psi^\dagger(x,t)$) in linear order. This linear stability analysis corresponds to the Bogoliubov theory of the BEC system \cite{castin}, and also have an analogy in optomechanics, and also in other nonlinear systems, especially in hydrodynamics. Inserting the separation of the field operators \eqref{eqs:sepmf} into Eqs.\ \eqref{eq:eqmores} and \eqref{eq:eqmoat}, and neglecting fluctuations higher than first order, one arrives to
\begin{subequations}
\label{eqs:fluct}
\begin{widetext}
\begin{align}
i\frac{\ud}{\ud t}\delta a(t)&=\Big[-\DC+N\langle U\rangle - i\kappa\Big]\delta a(t)
+N\alpha_{\text{ss}}\int \varphi(x) U(x)\Big[\delta \tilde\Psi(x,t)+\delta\tilde\Psi^\dagger(x,t)\Big]\,dx+i\xi(t),\label{eq:fluctrad}\\
i\hslash\frac{\partial}{\partial t}\delta\tilde\Psi(x,t)&=\bigg[\frac{\hslash^2\Delta}{2m} + V_{\text{ext}}(x)+\hslash|\alpha_{\text{ss}}|^2 U(x)-\mu+gN\varphi^2(x)\bigg]\delta\tilde\Psi(x,t)+\hslash U(x)\varphi(x)\Big[\alpha^*_{\text{ss}}\delta a(t)+\alpha_{\text{ss}}\delta a^\dagger(t)\Big]\nonumber\\
&+gN\varphi^2(x)\Big[\delta\tilde\Psi^\dagger(x,t)+\delta\tilde\Psi(x,t)\Big].\label{eq:fluctat}
\end{align}
\end{widetext}
\end{subequations}
where the zeroth order terms cancel, since they fulfill Eq.\ \eqref{eq:gpeat} and \eqref{eq:adiabel} with $\ud\alpha_{\text{ss}}/\ud t=0$. We have introduced $\delta\tilde\Psi(x,t)=N^{-1/2}\delta\Psi(x,t)\,e^{i\mu t/\hslash}$.

A closer look on Eqs.\ \eqref{eqs:fluct} reveals that the time evolution of the annihilation operators couple to those of the creation operators. It is a consequence of the complex nature of the photonic and particle fields. In order to diagonalize Eqs.\ \eqref{eqs:fluct} one can choose two equivalent ways. Either one can separate the complex quantities into real and imaginary parts and study their time evolution; this is the approach mainly used in optomechanical studies. The other way is to diagonalize the set of equations not just containing $\delta a$ and $\delta\tilde\Psi$, but also their hermitian adjoints $\delta a^\dagger$ and $\delta\tilde\Psi^\dagger$; this approach is familiar from the Bogoliubov-de Gennes theory of superfluidity. 

Adopting the Bogoliubov-de Gennes way we gather the fluctuations into the following column vector $R=(\delta a, \delta a^\dagger, \delta\tilde\Psi, \delta\tilde\Psi^\dagger)^T$, where the superscript $T$ stands for transposition, and noises to the other column vector $Z=(\xi,\xi^\dagger,0,0)^T$. Eqs.\ \eqref{eqs:fluct}
 can now be cast into the closed form
\begin{subequations}
 \begin{equation} 
 \label{eq:eqmomat}
 i\frac{\partial}{\partial t}R(t)=\mat{M}\,R(t) + i Z(t),
\end{equation}
with
\begin{widetext}
\begin{align}
\mat{M}&=\left[
\begin{array}{c c c c}
A & 0 & N\alpha_{\text{ss}}X& N\alpha_{\text{ss}}X\\
0 & -A^* & -N\alpha_{\text{ss}}^*X& -N\alpha_{\text{ss}}^*X\\
\alpha_{\text{ss}}^*Y(x)& \alpha_{\text{ss}} Y(x)&\hslash^{-1}[H_0+gN\varphi^2(x)]&\hslash^{-1}gN\varphi^2(x)\\
-\alpha_{\text{ss}}^*Y(x)& -\alpha_{\text{ss}} Y(x)&-\hslash^{-1}gN\varphi^2(x)&-\hslash^{-1}[H_0+gN\varphi^2(x)]
\end{array}\right],\label{eq:matm}\\
H_0&=-\frac{\hslash^2\Delta}{2m}+V_{\text{ext}}(x)+\hslash|\alpha_{\text{ss}}(x)|^2U(x)-\mu+gN\varphi^2(x),\\
X\cdot f(x)&=\int\varphi(x)U(x)f(x)dx,\\
Y(x)&=U(x)\varphi(x),\\
A&=-\DC+N\langle U\rangle-i\kappa.
\end{align}
\end{widetext}
\end{subequations} 

Since $\delta a$ is not independent of $\delta a^\dagger$ and similarly $\delta\tilde\Psi$ is also not independent of $\delta\tilde\Psi^\dagger$, the matrix $\mat{M}$ has an important symmetry property. It is a consequence that the effect of hermitian conjugation of $R$ can be obtained with a linear transformation $\mat{C}$ that swaps the first row with the second one and simultaneously the third one with the fourth, so $R^\dagger=\mat{C}\,R$. It directly follows from Eq.\ \eqref{eq:eqmomat} and from this symmetry property that  
\begin{equation}
\label{eq:symprop}
\mat{C}\cdot \mat{M}\cdot \mat{C}=-\mat{M}^*.
\end{equation}

In order to study the correlations of the fluctuations one has to determine the time evolution of the fluctuation operators first.  One has to introduce quasi-normal modes that diagonalize Eq.\ \eqref{eq:eqmomat} and therefore have a simple time evolution. Let us denote by $r^{(k)}$ the right eigenvectors of $\mat{M}$, i.e.,
\begin{equation}
\label{eq:eigenprob}
\mat{M}\, r^{(k)}=\omega_k\,r^{(k)},
\end{equation} 
with $\omega_k$ being the  corresponding  eigenvalue of $\mat{M}$. The fluctuation operator $R$ can be expanded with the help of the eigenvectors (if they form a complete set)
\begin{equation} 
\label{eq:normexp}
R(t)=\sum_k \rho_k(t) r^{(k)},
\end{equation}
with $\rho_k$ being the operator, or expansion coefficient, of the quasi-normal mode $k$. The operator $\rho_k$ is given by:
\begin{equation} 
\rho_k(t)=\big(l^{(k)},R(t)\big),
\end{equation}
with $(\cdot,\cdot)$ is the Euclidean scalar product, and $l^{(k)}$ is the left eigenvector of $\mat{M}$, defined as
\begin{equation} 
\mat{M}^\dagger\,l^{(k)}=\omega_k^*\,l^{(k)}.
\end{equation}
The left and right eigenvectors are normalized as usual: $\big(l^{(k)}, r^{(l)}\big)=\delta_{k,l}$. With the help of Eqs.\ \eqref{eq:eigenprob} and \eqref{eq:normexp}, the normal modes obey the following uncoupled equation of motion:
\begin{equation} 
\label{eq:eqmonorm}
i\frac{\ud}{\ud t}\rho_k(t)=\omega_k\,\rho_k(t)+iQ_k(t),
\end{equation}
with the transformed noise operator $Q_k=\big(l^{(k)},Z\big)$. On integrating Eq.\ \eqref{eq:eqmonorm}, the time dependence of the normal mode operators can be obtained:
\begin{equation}
\label{eq:rhoint}
\rho_k(t)=e^{-i\omega_k t}\,\rho_k(0)+\int_0^t e^{-i\omega_k(t-t')}\,Q_k(t')dt'.
\end{equation}
For a dynamically stable sytem, one needs to have eigenvalues with negative imaginary parts. In this case fluctuations (the normal modes) decay to a steady-state. 

The symmetry property \eqref{eq:symprop} has an important consequence on the spectrum of $\mat{M}$. Namely, if $\omega=\epsilon-i\gamma$ is an eigenvalue of $\mat{M}$, than $-\omega^*=-\epsilon-i\gamma$ is also an eigenvalue of $\mat{M}$. The modes corresponding to these eigenvalues form a pair with positive and negative energies.

There is an important issue concerning the stability of the normal modes: not all of them include the radiation components. For simplicity, assume that the external potential is even with respect to the center of the cavity. In this case the combined external and optical potentials is also even, and parity is a symmetry of the full system. In this case the condensate wavefunction is also even; and the matrix $\mat{M}$ commutes with the parity operator. The eigenfunctions of $\mat{M}$ can therefore be classified by their symmetry (being odd or even). If the condensate fluctuation parts (third and fourth components) of an eigenvector $r^{(k)}$ are odd, than the result of the operator $X$ acting on these components of $r^{(k)}$ is zero, since $X$ contains an integration on the whole cavity axis and its kernel is even. For these modes there is no coupling term between the cavity field and the condensate part [see the first two rows of the matrix $\mat{M}$ in \eqref{eq:matm}], and, since damping of the modes comes from the cavity decay, these modes remain undamped and just marginally stable. Consequently, if we want to describe the steady-state values of the correlations of fluctuations we have to omit those normal modes which do not couple to the radiation field (assuming that they are initially not populated).      

\subsection{Bloch states, the effects of s-wave scattering and that of the collective coupling}

For further analysis we suppose that $V_{\text{ext}}(x)\equiv0$, i.e. the only potential in Eq.\ \eqref{eq:eqmoat} is the periodic optical potential induced by the resonator field. In this case the Bloch functions are good candidates for the complete set of single particle wavefunctions:
\begin{equation} 
\label{eq:blochfn}
\psi\nq(x)=\mathcal{N}e^{iqx}u\nq(x),
\end{equation}
with $u_{n,q}(x)$ being a periodic function of period $\lambda/2$. Here $n$ is the so-called band index and $q\in[0,4\pi/\lambda]$ is the quasi-momentum of the particle. $\mathcal{N}$ is the constant for normalization. If we impose the Born-von Karman boundary condition with $p$ periods (the quantization volume is $p\,L$), then the normalization constant $\mathcal{N}=p^{-1/2}$, and the functions $u\nq(x)$ are normalized to unity inside a period $L=\lambda/2$. For practical purposes, we also use plane-waves as basis of the $u\nq(x)$ functions (not depending on the quasi-momentum $q$)
\begin{equation} 
\label{eq:upw}
u\nq(x)=\frac{1}{\sqrt{L}}e^{ink_0x},
\end{equation}
with $k_0=2k=4\pi/\lambda$. (Of course the basis \eqref{eq:upw} is suitable to express all other $L$-periodic functions as Fourier series.)

The field operator of the atoms are expanded as
\begin{equation} 
\label{eq:fieldoppw}
\Psi(x)=\frac{1}{\sqrt{L\,p}}\sum\nq b\nq\,e^{i(nk_0+q)x}. 
\end{equation}
with $b\nq$ the annihilation operator of the plane-wave state $\psi\nq$. Here $n$ goes over all integers and $q=4\pi m/(\lambda p)$, with $m\in\lbrace0,1,\ldots p-1\rbrace$.

The equation of motion of the annihilation operators $b\nq$ is obtained easily from Eq.\ \eqref{eq:eqmoat}, by inserting the field operator \eqref{eq:fieldoppw} and using the orthogonality of the plane-wave states,
\begin{widetext} 
\begin{equation} 
i\hslash\frac{\ud}{\ud t}b\nq=\frac{\hslash^2(nk_0+q)^2}{2m}b\nq+\hslash\,a^\dagger a\,U_0\sum_{n'}L_{n,n'}b_{n'q}+\frac{2g}{\lambda p}\sum_{\stackrel{n_1n_2n_3}{q_1q_2q_3}}b^\dagger_{n_1q_1}b_{n_2q_2}b_{n_3q_3}\delta_{(n+n_1)k_0+(q+q_1),(n_2+n_3)k_0+(q_2+q_3)}.
\end{equation}
\end{widetext}
The first term corresponds to the kinetic energy, which is diagonal for the plane wave states used. 
The matrix $L_{n,n'}$ appearing in the optical potential is a simple, tridiagonal matrix
\begin{equation} 
L_{n,n'}=\frac{1}{4}\Big[2\delta_{n,n'}+\delta_{n,n'+1}+\delta_{n,n'-1}\Big].
\end{equation}
It is trivial that the kinetic energy is the lowest for $n=0, q=0$. The appearance of the optical potential makes the Hamiltonian nondiagonal in plane-wave basis, however, it mixes only operators with the same quasi-momentum $q$. Coupled plane waves can have indices $n$ differing only by $\Delta n=\pm1$. On the other hand, the s-wave scattering mixes operators with different $n$ and different $q$ momenta. The total momentum is conserved by the $\delta$ function imposing $(n+n_1-n_2-n_3)k_0+(q+q_1-q_2-q_3)=0$, which allows for normal scattering, when $q+q_1=q_2+q_3$, and $n+n_1=n_2+n_3$, and also for umklapp scattering processes, where the difference of the quasi momentum in the scattering is equal to a reciprocal lattice vector: $q+q_1=q_2+q_3-\Delta n\cdot k_0$, and $n+n_1=n_2+n_3+\Delta n$. It is useful at this point to estimate the characteristic frequency corresponding to the atom-atom s-wave scattering based on the physical parameters relevant to the experimental situation in eg. Refs.~\cite{brennecke08a,ritter09a}. The s-wave scattering length for the $\left|1,-1\right\rangle$ states of the $^{87}\mathrm{Rb}$ atoms is about $5.3~\mathrm{nm}$. Assuming a particle number of $N=6\times10^4$ atoms distributed in the cavity lattice of period $L=\lambda/2=390~\mathrm{nm}$ and the cavity egg-crate potential containing $p=460$ valleys, corresponding to a cavity length of $180~\mu\mathrm{m}$, and the waist of the optical potential as $w=25~\mu\mathrm{m}$, the characteristic frequency of s-wave interaction can be around $\omega_{\text{sw}}=4\pi\hslash a N/(Lpw^2m)\approx2\pi\times4~\mathrm{Hz}$. (Here we have neglected any external trapping potential.) This means that the characteristic energy of s-wave scattering is 3 orders of magnitude lower than that of the recoil energy separating the bands from each other. As a result s-wave scattering dominantly occurs only with atoms in the same band and umklapp processes can be neglected. When incorporating the effects of the confining parabolic optical potential in Ref.~\cite{brennecke08a}, the condensate density grows roughly by a factor of 100, and results in the ratio $\omega_{\text{sw}}/\oR\sim 0.1$. In this situation the contribution of umklapp processes is still small enough to expect that it does not significantly change the forthcoming results. 

The mean-field equation for the condensate is obtained by the substitution
\begin{equation} 
\label{eq:mfpw}
b\nq=\sqrt{N}\delta_{q,0}\beta_n(t)+\delta b\nq(t),
\end{equation}
with $\beta_n$ the mean value of the annihilation operators. The $\delta_{q,0}$ condition for the mean-field comes from the fact that the ground state of the translationally invariant system is also invariant under discrete translation, i.e. the condensate wavefunction is periodic also with $L=\lambda/2$. (None of the terms in Eq.\ \eqref{eq:mfpw} violates the discrete translation of the system, and having $q\neq0$ would cost in kinetic energy.) The mean-field equation reads then 
\begin{multline} 
\label{eq:eqmomfpw}
i\hslash\frac{\ud}{\ud t}\beta_n(t)=\frac{\hslash^2n^2k_0^2}{2m}\beta_n(t)+\hslash|\alpha_{\text{ss}}|^2U_0\sum_{n'} L_{n,n'}\beta_{n'}(t)\\
+\frac{2g}{\lambda p}\sum_{n_1n_2n_3}\beta^*_{n_1}(t)\beta_{n_2}(t)\beta_{n_3}(t)\delta_{n+n_1,n_2+n_3},
\end{multline} 
with $\alpha_{\text{ss}}$ given by Eq.\ \eqref{eq:adiabel}, but now with
\begin{equation} 
\langle U\rangle=U_0\sum_{n,n'}\beta_n^*(t)L_{n,n'}\beta_{n'}(t).
\end{equation}
The steady-state components of the condensate amplitudes $\beta_n(t)$ evolve as
\begin{equation}
\label{eq:cwtdbw}
\beta_n(t)=\beta_n\,e^{-i\mu t/\hslash},
\end{equation} 
with $\mu$ being the chemical potential. It can be seen from Eq.\ \eqref{eq:eqmomfpw} that $\beta_n$ can also be chosen real for all $n$. Since $L_{n,n'}=L_{-n,-n'}$, and due to the symmetry of Eq.\ \eqref{eq:eqmomfpw} it follows that, $\beta_n=\beta_{-n}$, which means, that the Fourier expansion of the condensate wavefunction contains only cosine terms.

The equation of motion for the fluctuations, in this representation, is given by
\begin{figure}[tb!]
\begin{center}
  \includegraphics[scale=0.7]{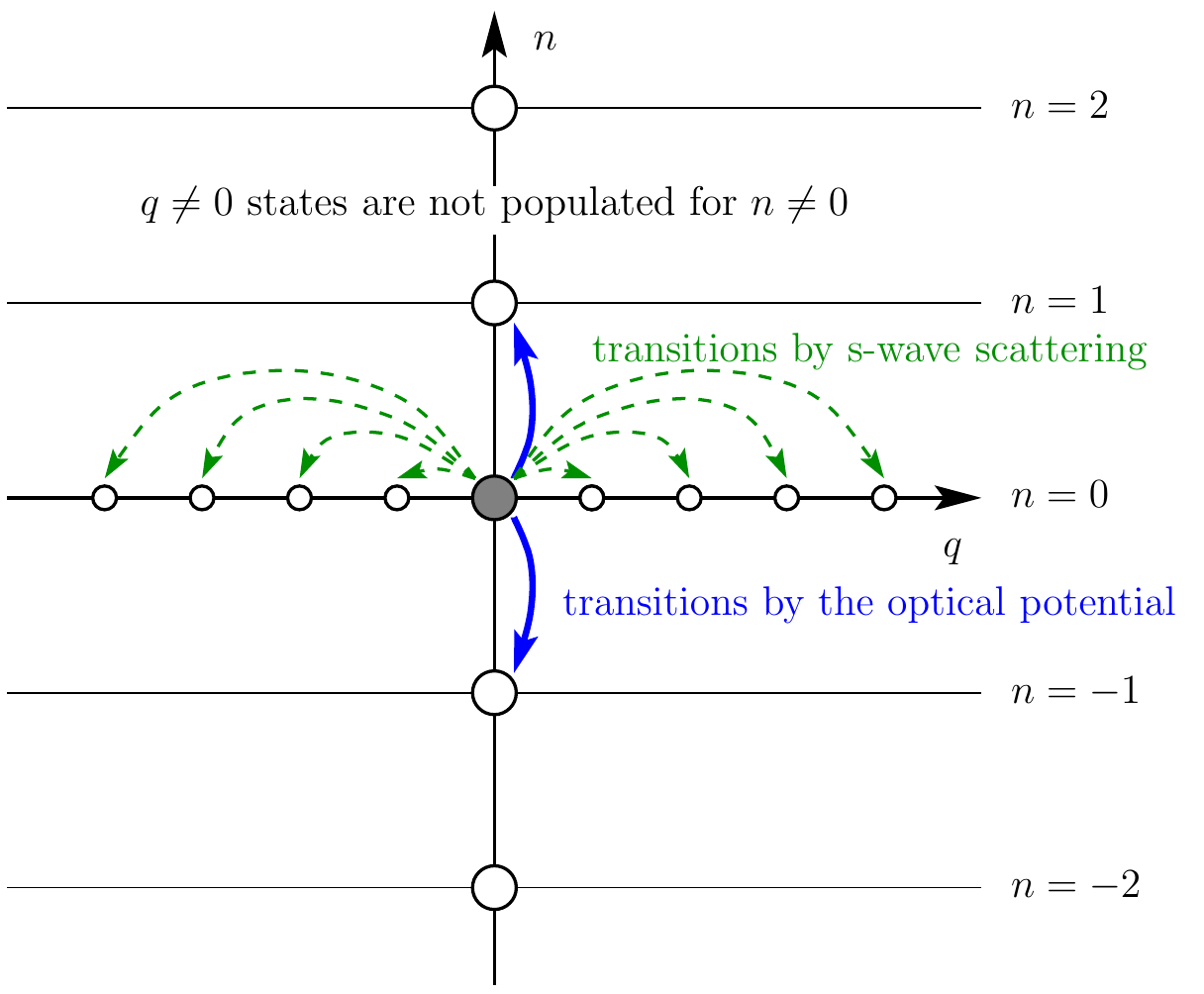}
  \caption{(color online) Schematic diagram of fluctuations coupling to the the homogeneous BEC in linear order. The grey circle represents the macroscopically occupied (BEC) state, while the empty circles represent states which are not macroscopically occupied. The arrows show how these states can be populated in linear order by scattering from the condensate via the interaction with the photon field and also via the s-wave scattering.}
  \label{fig:scattering}
\end{center}
\end{figure}
\begin{subequations}
\label{eqs:fluctbw} 
\begin{multline} 
i\frac{\ud}{\ud t}\delta a(t)=\Big[-\DC+N\langle U\rangle-i\kappa\Big]\delta a(t)\\+NU_0\alpha_{\text{ss}}\sum_{n,n'}\beta_n L_{n,n'}\Big(\delta \tilde b_{n',0}+\delta \tilde b_{-n',0}^\dagger)+i\xi(t),
\end{multline}
\begin{multline}
i\hslash\frac{\ud}{\ud t}\delta\tilde b\nq=(H_0)_{n,n'}\,\delta\tilde b_{n',q}\\
+\hslash U_0\delta_{q,0}(\alpha_{\text{ss}}^*\delta a+\alpha_{\text{ss}}\delta a^\dagger)\sum_{n'}L_{n,n'}\beta_{n'}\\
+\frac{2g}{\lambda p}\sum_{n_1,n_2,n'}\beta_{n_1}\beta_{n_2}\big(\delta\tilde b_{n',q}
+\delta\tilde b_{-n',-q}^\dagger\big)\delta_{n+n_1,n'+n_2},
\end{multline}
with
\begin{multline}
(H_0)_{n,n'}=\frac{\hslash^2(nk_0+q)^2}{2m}\delta_{n,n'}+\hslash|\alpha_{\text{ss}}|^2U_0L_{n,n'}\\
+\frac{g N}{L p}\sum_{n_1,n_2}\beta_{n_1}\beta_{n_2}\delta_{n+n_1,n'+n_2}-\mu\delta_{n,n'}.
\end{multline}
\end{subequations}
For a qualitative understanding of Eqs.~\eqref{eqs:fluctbw}, let us consider first the weak photon-atom coupling case, when either the photon number inside the cavity is small or $U_0$ is small. In this situation the atomic distribution is essentially unmodified by the periodic potential and the condensate wavefunction is almost homogeneous, i.e. $\beta_n=\delta_{n,0}\beta_0$. Fluctuations in the atomic density are induced from the homogeneous condensate. In linear order, the atom-light interaction excites fluctuations to bands $n\pm1$ with $q=0$, while s-wave interaction populates fluctuations with arbitrary $n$ and $q$. However, as discussed earlier, in the limit $\omega_{\text{sw}}\ll\oR$, s-wave scattering induces essentially only intra-band transitions, and therefore, for a homogeneous condensate, only states with $n=0$ are populated by s-wave interaction. Figure \ref{fig:scattering} depicts graphically the population of fluctuations excited from the homogeneous condensate wavefunction in leading order. Moreover, in experiments \cite{brennecke08a,ritter09a} the one atom light shift is a tunable parameter, it's typical value is chosen to be on the order of the recoil frequency,  i.e. $U_0\approx2\pi\times 4~\mathrm{kHz}$ and is also much larger than the characteristic frequency of s-wave scattering. Therefore on the time scale in which the steady-state is reached, the effect of s-wave scattering is still not significant. When describing the steady state, one can safely neglect it as a first approximation and consider each band represented by a single state vector with $q=0$.

Note that there is an interesting complementary regime in which the relevant states are localized, Wannier-type ones formed by the coherent superposition of many quasi-momentum states with $q\neq 0$. To describe many body effects of the atomic degree of freedom in this regime, one can resort to a Bose-Hubbard model with self-consistent parameters \cite{maschler08,vukics07b,vukics09,larson07a,larson07b,zoubi09a}, in which collisions play a vital role but interband transitions are usually neglected. 
%Works focusing on the many body effects of the atomic degree of freedom \cite{larson07a,larson07b,zoubi09a} incorporate atom atom scattering in a convenient and consistent way by a dynamical lattice treatment, where the parameters of the applied Bose-Hubbard model are to be determined self-consistently. In theses approaches the operator nature of the radiation field is lost, and so its correlations with the atoms. In contrast, in recent papers \cite{zhang09a,goldbaum10a} the effect of s-wave scattering has been taken into account by considering a narrow distribution of quasimomentum of the condensate wavefunction around each reciprocal lattice vector, however, a simple and consistent incorporation of atom-atom scattering rendering the atom-field correlation tractable remains still an open issue.

\subsection{Optomechanics}

In the experimental situation of Refs. \cite{brennecke08a,ritter09a} the BEC wavefunction can be considered almost homogeneous, with a condensate fraction in the $\beta_0$ state containing $6\times10^4$ atoms, while the next state with $n=\pm1$, i.e. the atomic motional state with wavefunction $\cos k_0 x$ containing only a few hundred atoms. Let us therefore restrict the Hilbert space of the one-particle atomic motion into this relevant 2-dimensional subspace, containing the homogeneous single particle wavefunction and that of $\cos k_0 x$. The atomic field operator becomes
\begin{equation} 
\label{eq:fopom}
\Psi(x)=\varphi_0(x) b_0 + \varphi_1(x)b_1 ,
\end{equation} 
with the single-particle wavefunctions $\varphi_0(x)=L^{-1/2}$, and $\varphi_1(x)=\sqrt{2/L}\cos(k_0 x)$, and their corresponding annihilation operators $b_0$ and $b_1$. The single-particle wavefunction $\sin(k_0x)$, corresponding to the antisymmetric combination of the $n=\pm1$ states, is omitted, since this wavefunction doesn't couple to the photonic field due to symmetry reasons mentioned above.

In this model the atomic motion is represented in a two-mode Fock space. The mean field expansion is 
\begin{equation} 
\label{eq:fodecom}
b_n=\sqrt{N}\beta_n+\delta b_n,\quad n=0, 1;
\end{equation}
where $\beta_n$ is again the representation of the condensate wavefunction in the 2-dimensional Hilbert space, normalized to unity. The presence of a Bose-Einstein condensate distinguishes a subspace of  one single mode. Accordingly, the fluctuation operators can be expanded to a part parallel to the condensate and to one orthogonal to it:
\begin{equation} 
\label{eq:parandorthomode}
\delta b_n= \beta_n\delta b + \gamma_n\delta c,
\end{equation}
with $\gamma_n$ the unit vector orthogonal to the condensate: $\gamma_n=(-\beta_1,\beta_0)^T$. The part, $\delta b$, parallel to the condensate can be related to the arbitrariness of the phase of the condensate, and corresponds to a zero mode. The orthogonal part, $\delta c$ is the sole degree of freedom. Therefore the fluctuations of the atomic field, similarly to the photon field, has a single degree of freedom. The fluctuations of the combined atom-resonator field system has 2 degrees of freedom, and is analogous in many ways to cavities with a moving mirror, to the so called optomechanical systems.

In order to obtain the mean-field equations and the fluctuation equations in linear order one can start from Eq.\ \eqref{eqs:eqmo} and use the truncated field operator \eqref{eq:fopom} and then the mean-field ansatz \eqref{eq:fodecom}. The mean-field equations for the condensate now read
\begin{equation} 
\label{eq:gpeom}
i\frac{\ud}{\ud t}\beta_k(t)=\Big[4\,\oR\big(P_1\big)_{k,l}+|\alpha_{\text{ss}}|^2 U_{k,l}\Big]\beta_l(t),
\end{equation}
with $P_1=\mathrm{diag}(0,1)$, the projection matrix to the subspace of $\varphi_1(x)$; the matrix representing the light-shift is given by
\begin{equation} 
U_{k,l}=\langle\varphi_k|U(x)|\varphi_l\rangle=\frac{U_0}{2}\left[
\begin{array}{c c}
1&\sqrt{2}/2\\
\sqrt{2}/2&1
\end{array}\right],
\end{equation}
$\alpha_{\text{ss}}$ given by Eq.\ \eqref{eq:adiabel}, with $\langle U\rangle=\beta_k^*(t) U_{k,l} \beta_l(t)$. (Note the convention of automatic summation over repeated indices.) The condensate wavefunction also has the time dependence of Eq.\ \eqref{eq:cwtdbw}, with $\beta_k$ chosen real. Eq.\ \eqref{eq:gpeom} can also be solved either by imaginary time propagation or by direct algebraic means using Eq.\ \eqref{eq:cwtdbw} and the normalization condition: $\beta_0^2+\beta_1^2=1$.

The equations of motion for the fluctuation operators can be derived analogously to the way as Eqs.\ \eqref{eqs:fluctbw} was obtained. It is appropriate to separate the trivial time dependence due to the chemical potential via the definition $\delta\tilde b_k=N^{-1/2}\,e^{i\mu t/\hslash}\,\delta b_k$. With this
\begin{subequations}
\begin{align} 
i\frac{\ud}{\ud t}\delta\tilde a&=(-\delta - i\kappa)\delta a + N\alpha_{\text{ss}}\beta_k U_{k,l}(\delta\tilde b_l + \delta\tilde b_l^\dagger),\label{eq:eqmoaom}\\
i\frac{\ud}{\ud t}\delta\tilde b_k&=K_{k,l}\delta\tilde b_l + (\alpha_{\text{ss}}^*\delta a + \alpha_{\text{ss}}\delta a^\dagger)\,U_{k,l}\beta_l,\label{eq:eqmobom}
\end{align} 
with $\delta=\DC - N\langle U\rangle$, and
\begin{equation} 
\label{eq:gham}
K_{k,l}=4\oR \big(P_1\big)_{k,l}-\frac{\mu}{\hslash}\delta_{k,l}+|\alpha_{\text{ss}}|^2 U_{k,l}.
\end{equation}
\end{subequations}
The matrix $K$ can be thought of being the grand canonical Hamiltonian of the system, and by virtue of Eqs.\ \eqref{eq:gpeom} and \eqref{eq:cwtdbw} $K_{k,l}\,\beta_l=0$. Equations \eqref{eq:eqmoaom} and \eqref{eq:eqmobom} form a linear eigenvalue problem for the fluctuations. It is easy to check  that the trial function of $\delta a=0$ and $\delta\tilde b_k=\beta_k\delta \tilde b$ is a constant solution, i.e. it is a zero mode. It also follows from the normalization of $\beta_k$ that $\delta \tilde b$ is antihermitian. With the decomposition $\delta\tilde b_k=\beta_k\,\delta\tilde b + \gamma_k\,\delta\tilde c$ one can arrive at a closed set of equations between the fluctuations $\delta a$ and $\delta\tilde c$ and their hermitian adjoints.
\begin{subequations} 
\label{eqs:fluctom}
\begin{align}
i\frac{\ud}{\ud t}\delta\tilde a&=(-\delta - i\kappa)\delta a + N\alpha_{\text{ss}}(\beta_k U_{k,l}\gamma_l)(\delta\tilde c + \delta\tilde c^\dagger),\label{eq:eqmoaomf}\\
i\frac{\ud}{\ud t}\delta\tilde c&=(\gamma_kK_{k,l}\gamma_l)\,\delta\tilde c + (\gamma_k U_{k,l}\beta_l)\,(\alpha_{\text{ss}}^*\delta a + \alpha_{\text{ss}}\delta a^\dagger).\label{eq:eqmobomf}
\end{align}
\end{subequations}
Note, that in the optomechanical model s-wave scattering can not be included in general, because it populates states with $q\neq0$, which are disregarded in this model. It is a possibility to include some remaining effects of the atom-atom collision by coupling only the two states under consideration, however, it would result in a nonlocal and unphysical interaction in coordinate space. Or, in order to extend the optomechanical model, although still lacking the full consideration of s-wave scattering explicitly, the higher quasi-momentum states can be taken into account either by considering a (close to, but non-) plane wave condensate wavefunction \cite{zhang09a}, or in a two-fluid model \cite{goldbaum10a}.

\section{Results}
\label{sec:res}

In the following we summarize and discuss our results based on the numerical solutions of Eqs.\ \eqref{eqs:gpe} and \eqref{eqs:fluct}, and compare them with those of Eqs.\ \eqref{eq:gpeom} and \eqref{eqs:fluctom} and to the findings of the Esslinger group \cite{brennecke08a,ritter09a}.

\subsection{The mean-field solution}

\begin{figure}[tb!]
\begin{center}
  \includegraphics[scale=0.7]{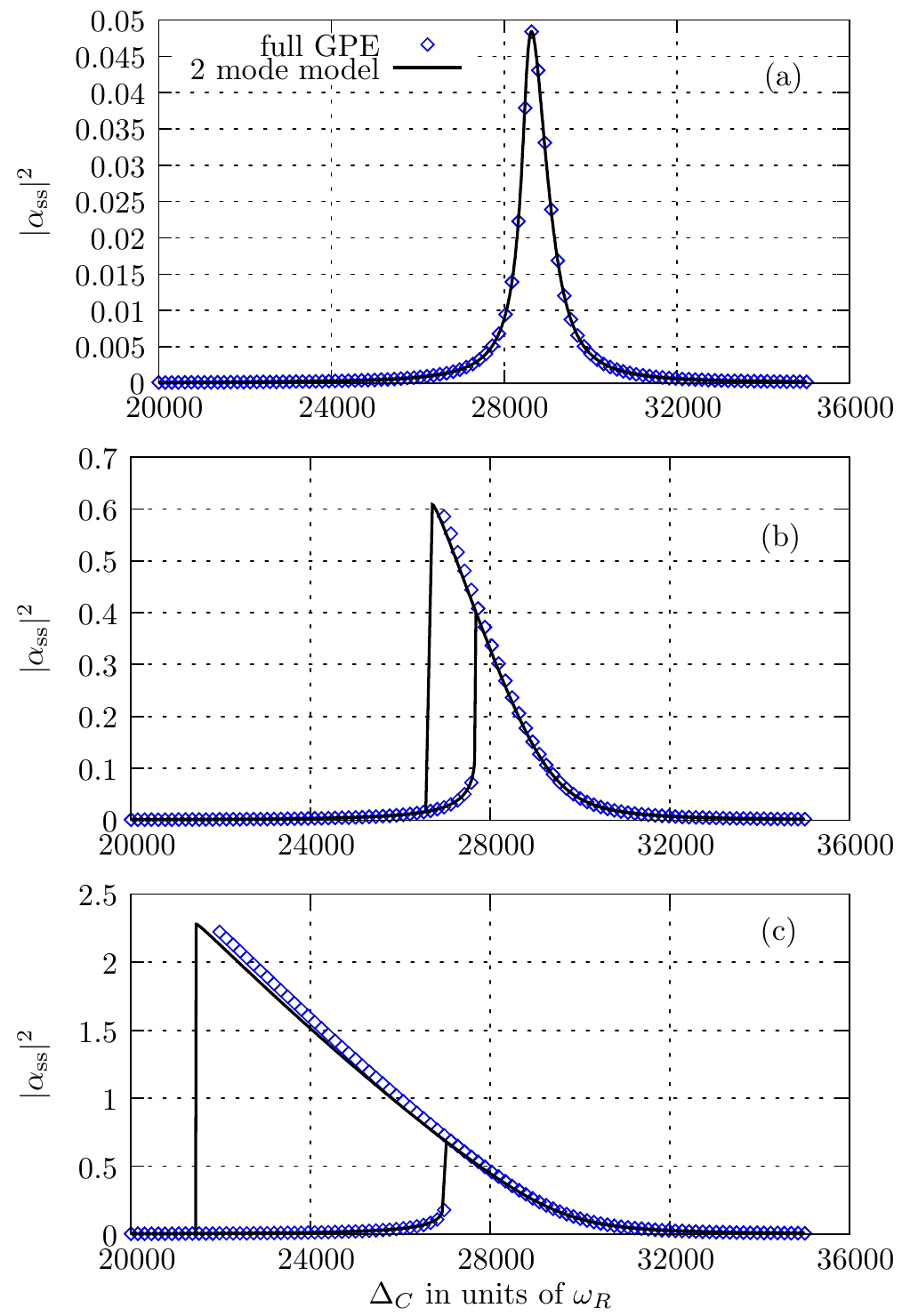}
  \caption{(color online) The mean cavity photon number $|\alpha_{\text{ss}}|^2$ as a function of the cavity detuning $\DC$. The parameters are: $N=6\times10^4$, $U_0=0.96\,\oR$, $\kappa=363.9\,\oR$. The pumping strength is different on the 3 panels: $\eta=(80.06, 283.8, 549.5)\oR$ for panels a, b, c, respectively.}
  \label{fig:phnum}
\end{center}
\end{figure}

For a comparison between the full GPE solution and the optomechanical model we numerically solve the Gross-Pitaevskii Equation \eqref{eq:gpeat} on a 200 point grid with imaginary time propagation with the steady-state value of the mean radiation field amplitude \eqref{eq:adiabel} and compare it to the solution of \eqref{eq:gpeom} (also in imaginary time). The results for the mean photon number are plotted in Fig. \ref{fig:phnum}.  The bistable behavior is nicely exhibited in these figures. In panel a) the pumping strength $\eta=80.06\,\oR$ is below a threshold value, the resonance curve gives a unique solution for all detunings $\DC$. Notice, that in Eqs.\ \eqref{eq:gperad}, or \eqref{eq:adiabel} the effective detuning of the resonator field is $\delta=\DC-N\langle U\rangle$, i.e., the light shift further detunes the cavity. On the left side of the resonance, the effective detuning $\delta$ is negative, the cavity is detuned to the red. We will see in the next subsection that this situation corresponds to cavity cooling, where the imaginary part of the fluctuation spectrum of the combined atom-photon system is negative (or nonpositive for the many mode system). Fluctuations decay to zero. On the right side of the resonance, where $\delta$ is positive, the imaginary part of the fluctuation spectrum changes its sign. Fluctuations grow exponentially on this side of the resonance, that is, the slightly retarded cavity field dynamics heats the atomic motion instead of cooling it. Therefore the solution above the resonance ($\delta>0$) is dynamically unstable. It is only due to the adiabatic elimination of the photon field that the numerical method finds this solution as well. 
Note, that the parameters of our calculations mimics those of Ref. \cite{ritter09a}. The agreement in the photon number with Fig. 3 of Ref. \cite{ritter09a} is well within the systematic uncertainty of the photon number estimation of the experiments (25\%). With the use of the recoil frequency for Rubidium atoms, $\oR=2\pi\times3.57~\mathrm{kHz}$, we can compare the location of the resonance points to the experimental findings, too. In panel (a), the resonance point is at around $28800\,\oR\approx2\pi\times103~\mathrm{MHz}$,  in panel (b) the two instability points are at around $26700\,\oR\approx2\pi\times95~\mathrm{MHz}$ and $27700\,\oR\approx2\pi\times99~\mathrm{MHz}$. All these are within 5\% of the respective experimental value.  In panel (c) the resonance points are at around $21500\,\oR\approx2\pi\times77~\mathrm{MHz}$ and $27000\,\oR\approx2\pi\times96~\mathrm{MHz}$. This latter is only 2\% higher than in the experiment, but the lower point is contrasted to the $2\pi\times84~\mathrm{MHz}$ value (about 10 \% deviation). In our case the bistable region is a bit wider than in the experiments. This discrepancy might be attributed mainly to the uncertainty in the effective detuning $\delta$ which we obtain by an estimation of the actual atom number and overlap between the cavity potential and the atomic cloud.

For $\eta>\eta_c(\DC,\kappa,U_0,N)$, the threshold value of the pumping strength depending on the detuning, the cavity loss, the light shift and also  the atom number, there is a region where three solutions exist for the photon number and also for the condensate wavefunction. Two of them are stable solutions for the numerical method we use. These solutions are plotted in panels b) and c) for $\eta=283.8\,\oR$ and $\eta=549.5\,\oR$, respectively. The unstable solution is not plotted in the figure. We note, however, that the upper one of the two plotted solutions corresponds to the $\delta>0$, cavity heating, situation. So this solution is also unstable, dynamically. But since this instability is related to photon dynamics (neglected at this level) one can still find this solution by integrating the GPE for the atoms in imaginary time.

\begin{figure}[t!]
\begin{center}
  \includegraphics[scale=0.7]{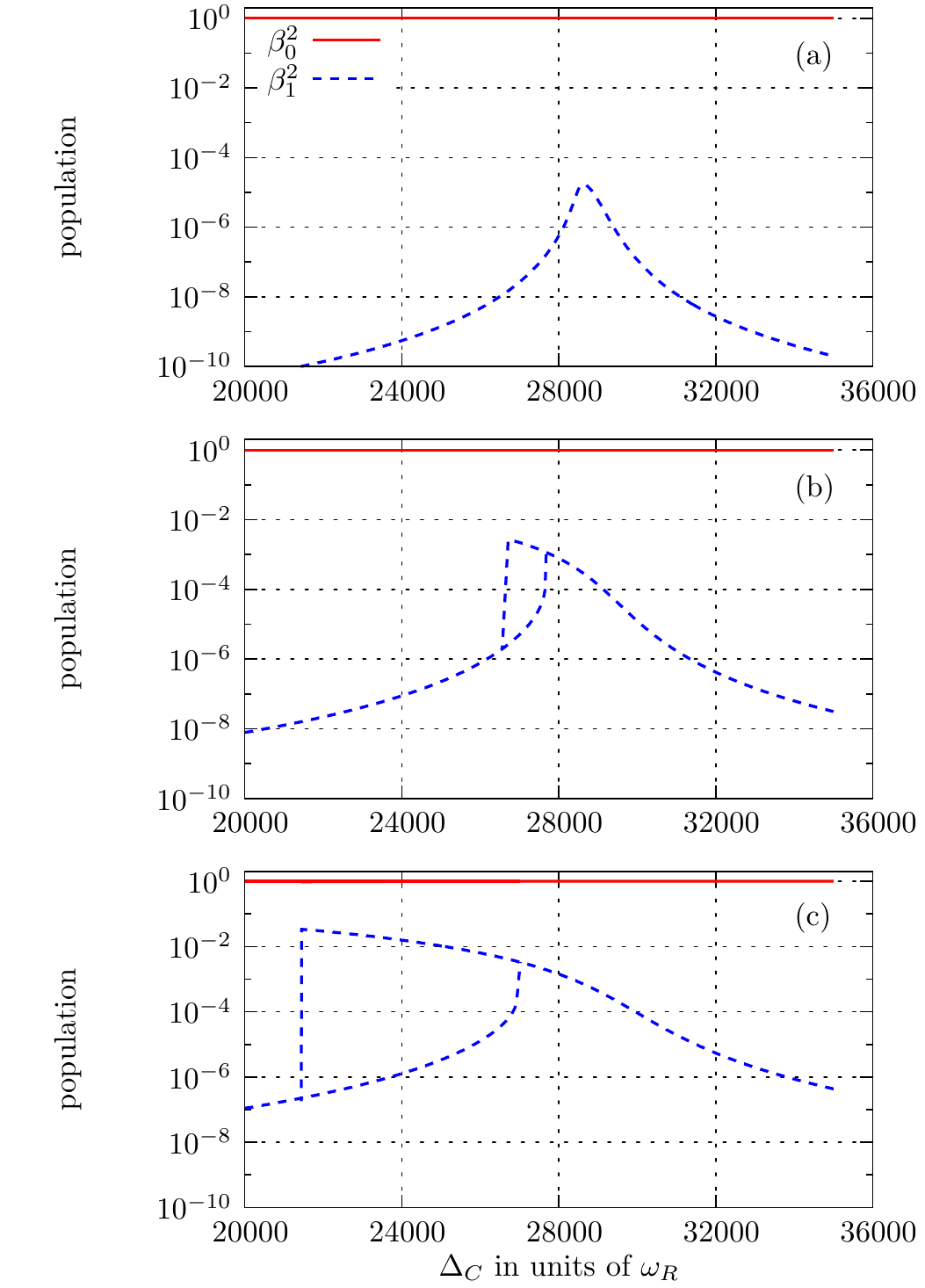}
  \caption{(color online) The squares of the condensate components $\beta_0^2$ and $\beta_1^2$ as a functions of the detuning $\DC$ in a semi-log scale. The parameter settings are the same as for Fig. \ref{fig:phnum}.}
  \label{fig:cond_comp}
\end{center}
\end{figure}
The fraction of condensate atoms occupying the homogeneous and the $\cos k_0 x$ states are plotted on Fig. \ref{fig:cond_comp}. The parameter settings are the same as for Fig. \ref{fig:phnum}. Both the resonance feature and the bistabily is exhibited in the condensate component of the $\cos k_0 x$ mode. The condensate is almost homogeneous for the whole range, $\beta_1^2\ll\beta_0^2\approx1$, and $\beta_1^2$ shows a very similar dependence as the photon number. It is remarkable that a very small change in the shape of the condensate wavefunction causes a drastic increase or fall of the photon number. This can be elucidated by the collective coupling of the atoms to the resonator field. The detuning $\delta=\DC-N\langle U\rangle$, appearing in the denominator of the steady-state value of the resonator field \eqref{eq:adiabel}, can change a lot even if the variation of $\langle U\rangle$ is small, because of the large number $N$ multiplying it.

\subsection{Fluctuation spectrum}

\begin{figure*}[t!]
\begin{center}
  \includegraphics[scale=0.7]{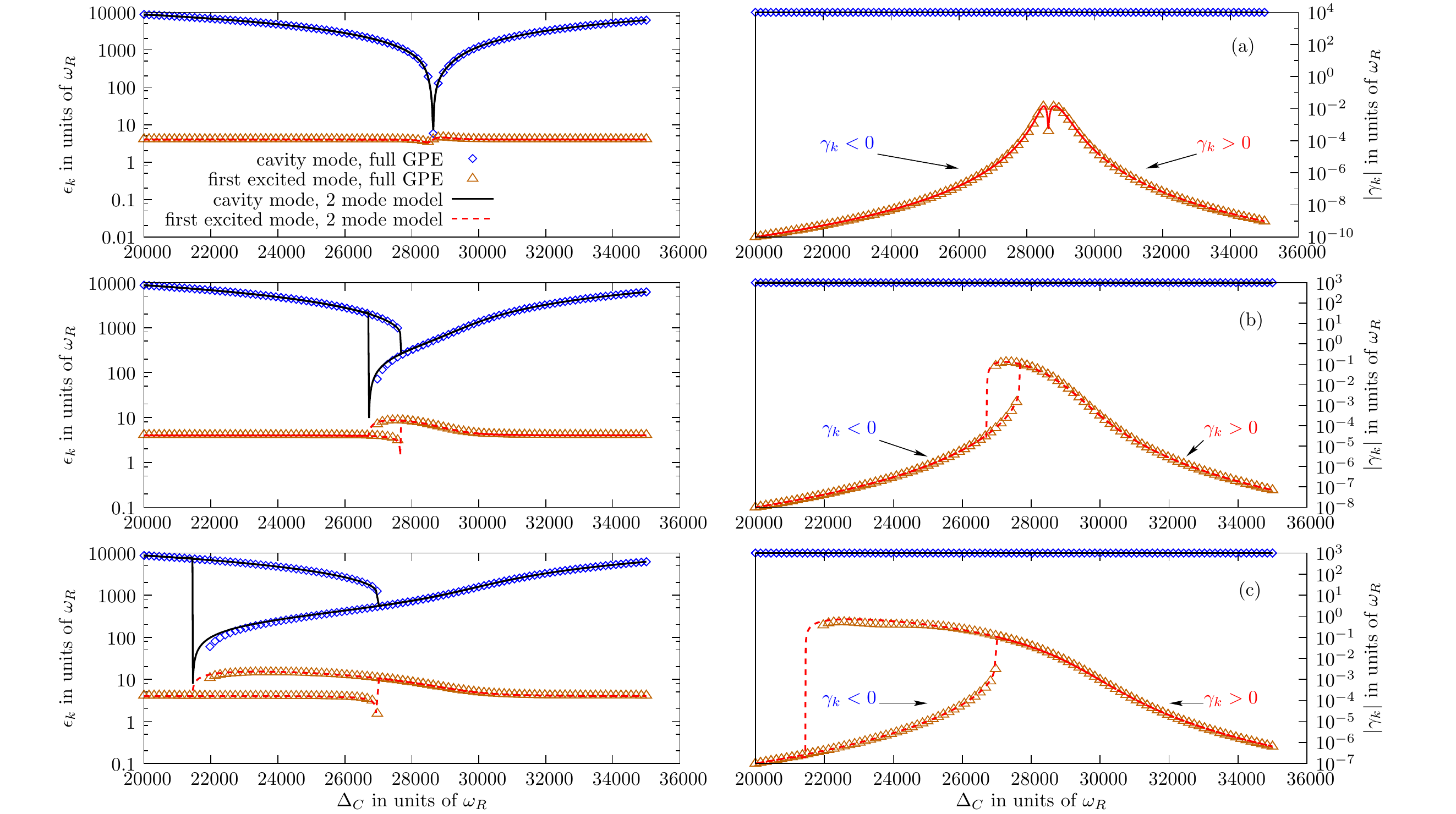}
  \caption{(color online) The fluctuation spectrum $\omega_k=\epsilon_k+i\gamma_k$ of the atom-cavity dynamics. The real and imaginary parts of the eigenvalues are plotted as functions of $\DC$. All quantities are measured in units of the recoil frequency. The parameter settings are the same as for Fig. \ref{fig:phnum}. Note, that the fluctuation spectrum of the full model, discretized on a 200 grid, contains 201 pairs of eigenvalues. We only plot those 2, which correspond to those of the optomechanical system.}
  \label{fig:fluctspect}
\end{center}
\end{figure*}
The linear stability analysis of the mean-field solution is done with the help of the linear equations \eqref{eqs:fluct} for the case of the full model, and with the help of Eqs.\ \eqref{eqs:fluctom} for the optomechanical approximation. Once the condensate wavefunction and the steady-state value of the mean cavity field is known, one can construct the matrix $\mat{M}$ and calculate its eigenvalues and eigenvectors. In the full model, discretized on a grid with 200 sites, $\mat{M}$ is a $402\times402$ matrix, since it acts on a row vector having $2$ elements for the photon fluctuation operator and its hermitian adjoint, and also having $400$ elements for the discretized field fluctuation operator and on its hermitian adjoint. For the optomechanical model one has a much smaller, $4\times4$ matrix, since the zero mode is already separated in this case, so one has $2$ components for the photon fluctuations and $2$ components for the atomic motion orthogonal to the condensate wavefunction. The numerical diagonalization was obtained with the help of the LAPACK package.

Note that $\mat{M}$ is a general complex matrix. Its eigenvalues are complex. However, due to the symmetry relation \eqref{eq:symprop} the eigenvalues of $\mat{M}$ come in pairs. Each element of the pair have the same imaginary part, while the real parts are just the opposite of each other. Figure \ref{fig:fluctspect} shows the eigenvalues of $\mat{M}$ for the same parameter settings as that of Figs. \ref{fig:phnum} and \ref{fig:cond_comp}. The left panels show the real part $\epsilon_k$ of the $k$th eigenvalue, while the right panels show the modulus of the imaginary parts $\gamma_k$ for $\omega_k=\epsilon_k+i\gamma_k$. Only those two eigenvalues are plotted, which are present in the optomechanical model and have positive real parts. The other two eigenvalues of the optomechanical model can be obtained simply by changing the sign of the real parts to negative: $\omega_{k}\rightarrow-\epsilon_k+i\gamma_k$. In the bistability regime, the excitation frequencies are presented for both mean field solutions. 

Again, all figures show that the optomechanical approximation very well reproduces the results of the full GPE simulation even for dynamical quantities. The two modes are easy to be physically interpreted. The first eigenvalue, having a bigger real part around $\delta=\Delta_C-N\langle U\rangle$ and an imaginary part almost exactly at $-\kappa$, corresponds to a mainly photonic mode. Note that the imaginary part of this mode remains also negative above the resonance. The other eigenvalue, having a real part around $4\,\oR$ corresponds mainly to atomic fluctuations. The imaginary part of this eigenvalue changes sign at resonance. Below resonance the imaginary part is negative (fluctuations in the atomic motion are damped due to the interaction with the resonator field \cite{horak01,gardiner:051603}), while above resonance the imaginary part becomes positive (fluctuations in the atomic motion are exponentially growing in time). The situation is clear on panel a), where the resonance point can be exactly defined. On panels b) and c) the coexistence of the two solutions makes the definition of the resonance point ambiguous. Nevertheless, on panels b) and c) the curves which continue to the left hand side of the figure correspond to the solutions with cavity cooling and those which continue to the right hand side correspond to the heating solution. The dynamical cooling and heating effects are closely related to the same effects at the single atom level \cite{horak97,maunz04,nussmann05a,vukics05a,vukics05b}.

In Refs. \cite{brennecke08a,ritter09a} the coherent atomic dynamics was also studied. The harmonic oscillator behavior of the low-energy atomic dynamics caused periodic oscillations in the output photon signal with a frequency close to $35~\mathrm{kHz}$, as reported and explained in Ref. \cite{brennecke08a}, or close to around $42~\mathrm{kHz}$ in Ref. \cite{ritter09a}. In our model the dynamics of the two coupled harmonic oscillators of the resonator field and the atomic collective motion is described by Eqs. \eqref{eqs:fluctom}. These are linearized equations of motion around the steady-state configurations. For very small fluctuations the semiclassical time evolution can be interpreted as orbitals around the steady-state fixed points with frequencies plotted in the left panels of Fig. \ref{fig:fluctspect}. Due to the imaginary parts the trajectories spiral closer to (for fluctuations around the cooling solution) or farther from (for fluctuations around the heating solution) the corresponding fixed point. From Fig. \ref{fig:fluctspect} we can see, that the frequency of atomic motion dominated fluctuations is close to $4\,\oR$ for parameters not very close to the bistable region. In the bistable regime two steady-state solutions exist, and correspondingly there are two fixed points in the harmonic oscillator phase space. According to Fig. \ref{fig:fluctspect} the oscillation frequency for the cooling fixed point remains around $4\,\oR$ and softens close to the endpoint, while for the heating solution the oscillation frequency grows to around $10\,\oR$ before softening at its endpoint. This kind of renormalization of the atomic dynamics by the interaction with radiation is often referred to as the `optical spring effect'. The approximately $10\,\oR$ angular frequency quantitatively gives back the experimentally found $42~\mathrm{kHz}$ oscillation frequency of the Fig. 3C inset of Ref. \cite{ritter09a}.

It is interesting to point out that in the bistable region even small atomic fluctuations can result in a phase-space trajectory that orbits around both of the fixed points [see Fig. 3 of Ref. \cite{brennecke08a}]. In this case such a linearization strategy simply can not work because the transition between the two fixed points is necessarily a nonlinear effect. However, if the trajectory would lie in the basin of the linear region of the corresponding fixed points everywhere except the small region of the separatrix, one could expect the oscillation to be described by the above two frequencies: when the system is in a part of the trajectory inside the attraction basin of the cooling fixed point the angular frequency would be around $4\,\omega_R$, while after crossing the separatrix it would change to around $10\,\omega_R$. Such a trajectory of coherent oscillations would also cause a periodic output photon signal but both of the frequencies would appear, i.e. the count rate peaks would come in pairs. 

\begin{figure}[t!]
\begin{center}
  \includegraphics[scale=0.7]{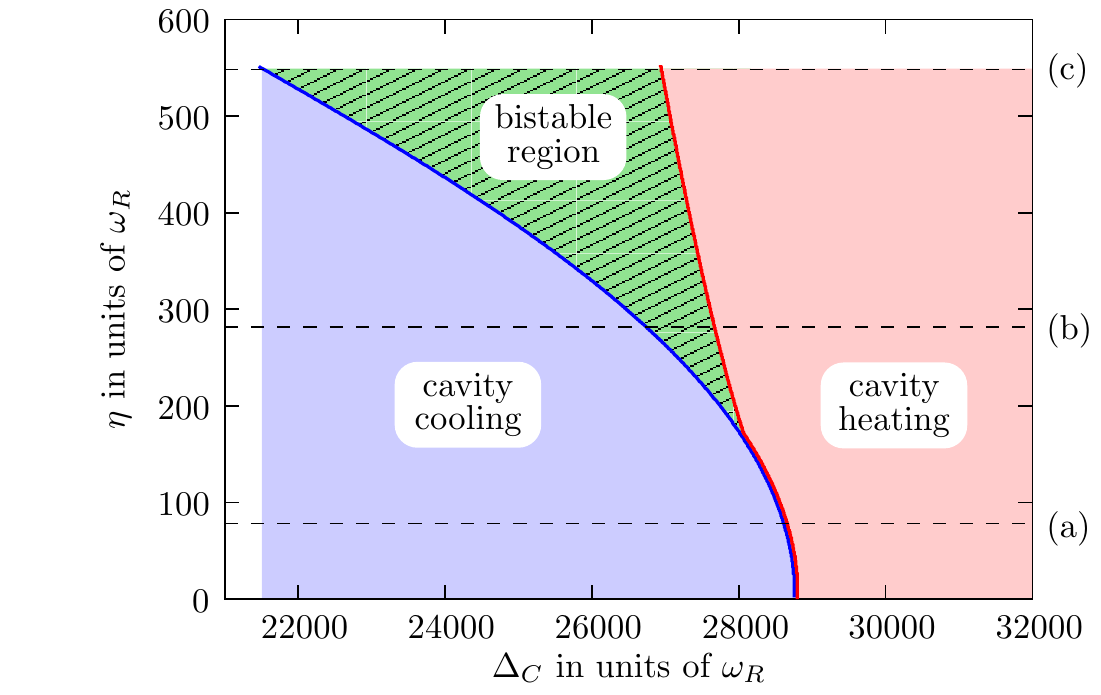}
  \caption{(color online) The nonequilibrium phase diagram of the system indicating the parameter regions with full dynamical stability (the cooling region) and the region with a dynamical instability caused by the energy transfer of the cavity (the heating region). The horizontal lines represent the parameter $\eta$ at values corresponding to the subfigures a) b) and c) of other figures with numeric results.}
  \label{fig:phdiag}
\end{center}
\end{figure}
Figure \ref{fig:phdiag} depicts the regions in the parameter space which represent the heating and (cooling) solutions, i.e. the solutions with dynamical instability (stability). The bistable region, where both a heating and a cooling solution can be found, is wedged between the two distinct regions. The tip of the wedge corresponds to the critical point $({\DC}_c,\eta_c)$.

\subsection{Correlations and entanglement}

Second order correlations of the fluctuations are calculated with the help of the quasi-normal modes and the noise correlations, since by virtue of Eqs.\ \eqref{eq:normexp} and \eqref{eq:rhoint}
\begin{multline} 
\label{eq:equtimecorr}
\langle R_k(t)\,R_l(t)\rangle=\sum_{m,n}\langle\rho_m(t)\,\rho_n(t)\rangle\,r^{(m)}_k\,r^{(n)}_l\\
=\sum_{m,n}r^{(m)}_k\,r^{(n)}_l\bigg[e^{-i(\omega_m+\omega_n)t}\langle\rho_m(0)\,\rho_n(0)\rangle\\
+\int_0^t dt_1 dt_2e^{-i\omega_m(t-t_1)}e^{-i\omega_n(t-t_2)}\langle Q_m(t_1)Q_n(t_2)\rangle\bigg].
\end{multline}
For a stable system, where all the eigenvalues have negative imaginary parts, the first term on the right hand side, corresponding to the initial condition of the fluctuations, vanishes for times much longer than the characteristic decay times of the system. For the second term one can use the noise correlation function \eqref{eq:noisecorr}, the definition of the vector noise $Z$ and that of $Q$. For $t\rightarrow\infty$ it follows straightforwardly, that
\begin{equation}
\label{eq:corrmat}
\langle R_k(t)\,R_l(t)\rangle\rightarrow2\kappa\sum_{m,n}\frac{l^{(m)*}_1\,l^{(n)*}_2\,r^{(m)}_k\,r^{(n)}_l}{i(\omega_m+\omega_n)},
\end{equation}
where we index the components of the row vectors $R$ and $Z$ starting from 1, and also dropped the exponential term vanishing for large times in the case of a stable system. Note that the zero mode, i.e., $\delta b$ in \eqref{eq:parandorthomode} representing the phase fluctuations of the condensate, do not contribute to this sum because its eigenfunction does not have photon component, i.e., $l^{(\text{zero mode})}_1=l^{(\text{zero mode})}_2=0$.

\begin{figure}[t]
\begin{center}
  \includegraphics[scale=0.7]{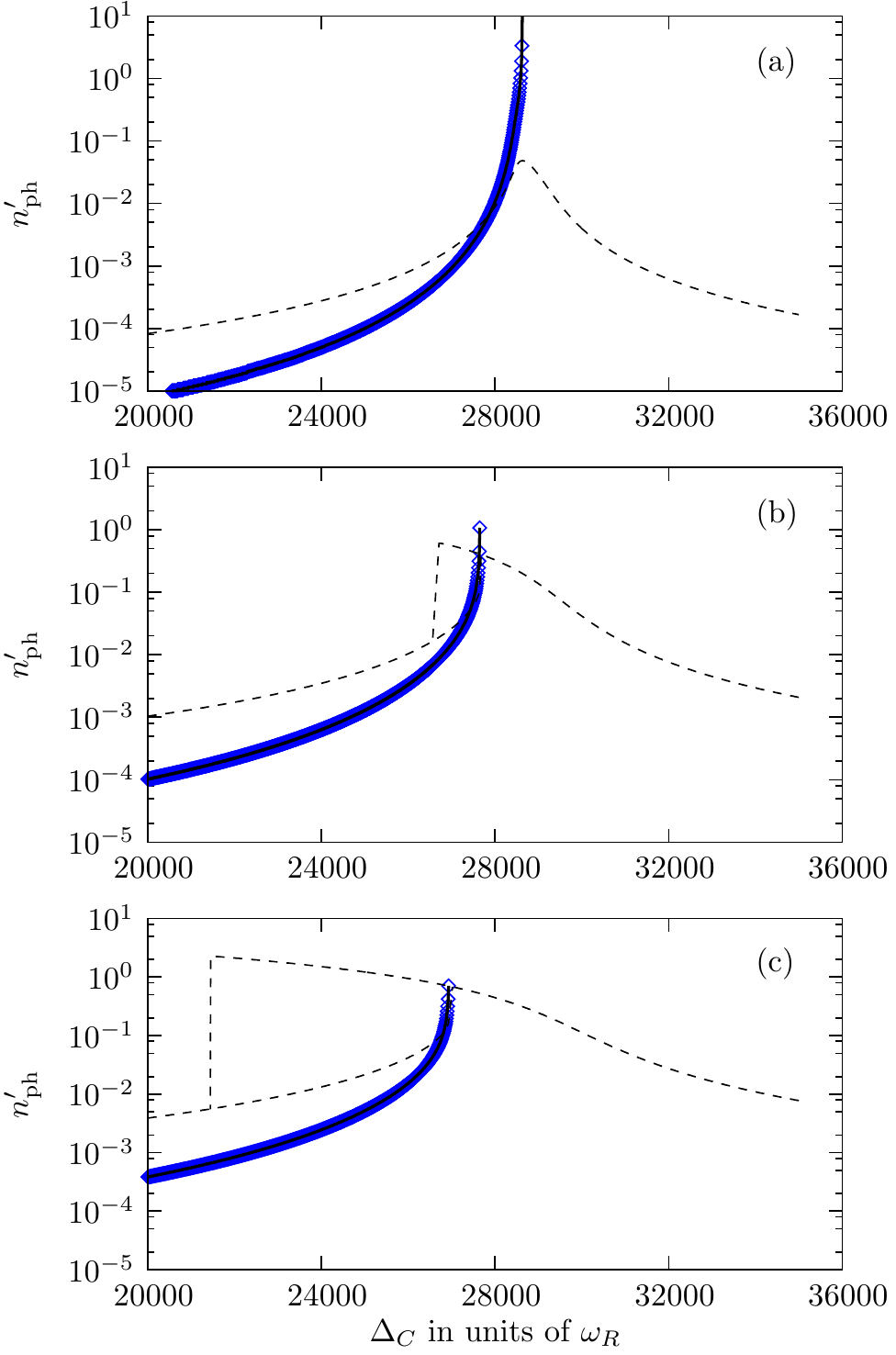}
  \caption{(color online) The nonclassical part of the photon number vs. the cavity detuning of the optomechanical model. For reference we have also plotted the mean-field solution, $|\alpha_{\text{ss}}|^2$, with a dashed line. All parameters are the same as for Fig. \ref{fig:phnum}.}
  \label{fig:phnumnonclass}
\end{center}
\end{figure}
\begin{figure}[tb]
\begin{center}
  \includegraphics[scale=0.7]{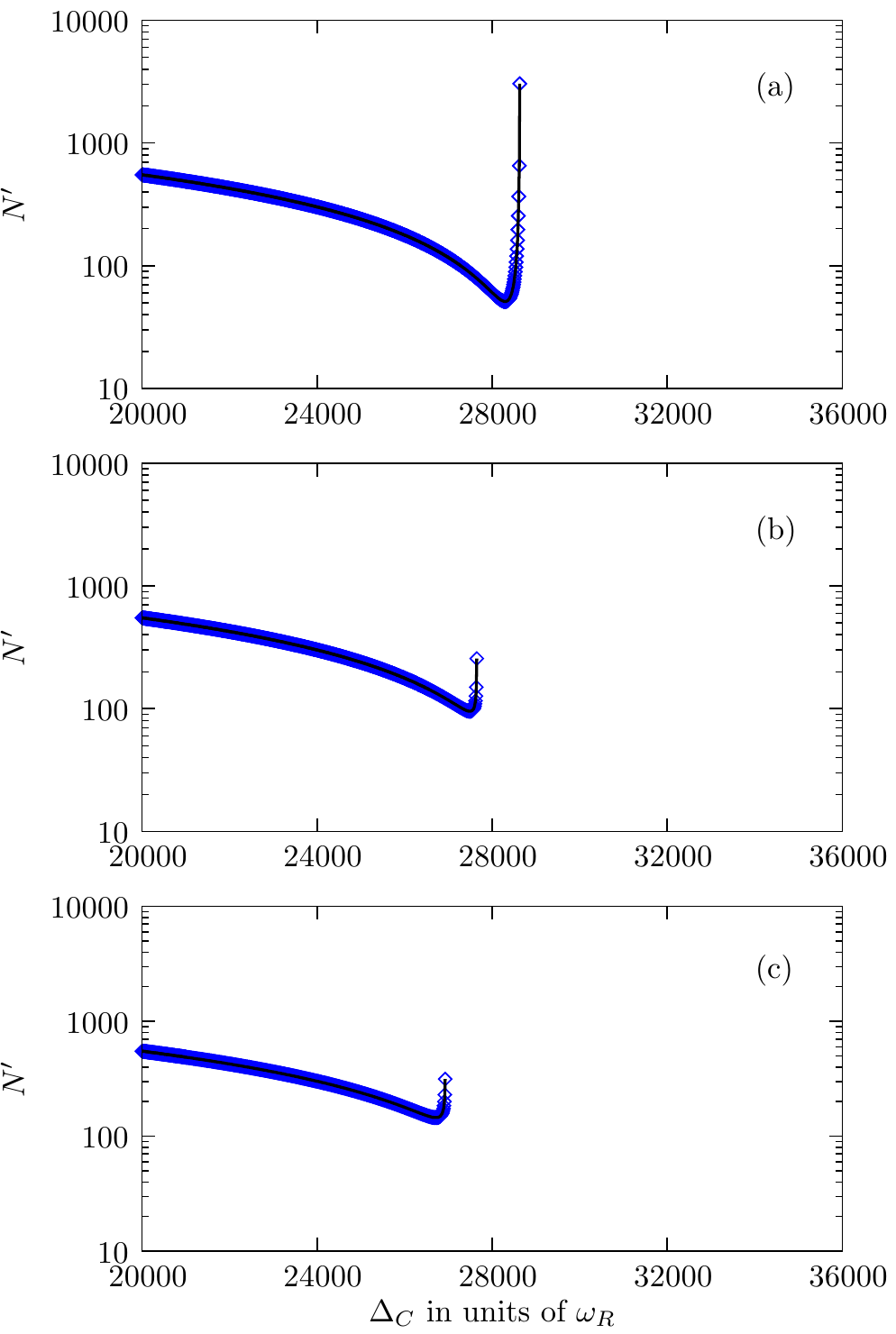}
  \caption{(color online) The depletion of the condensate vs. the cavity detuning of the optomechanical model. All parameters are the same as for Fig. \ref{fig:phnum}.}
  \label{fig:depletion}
\end{center}
\end{figure}
\begin{figure}[tb]
\begin{center}
  \includegraphics[scale=0.7]{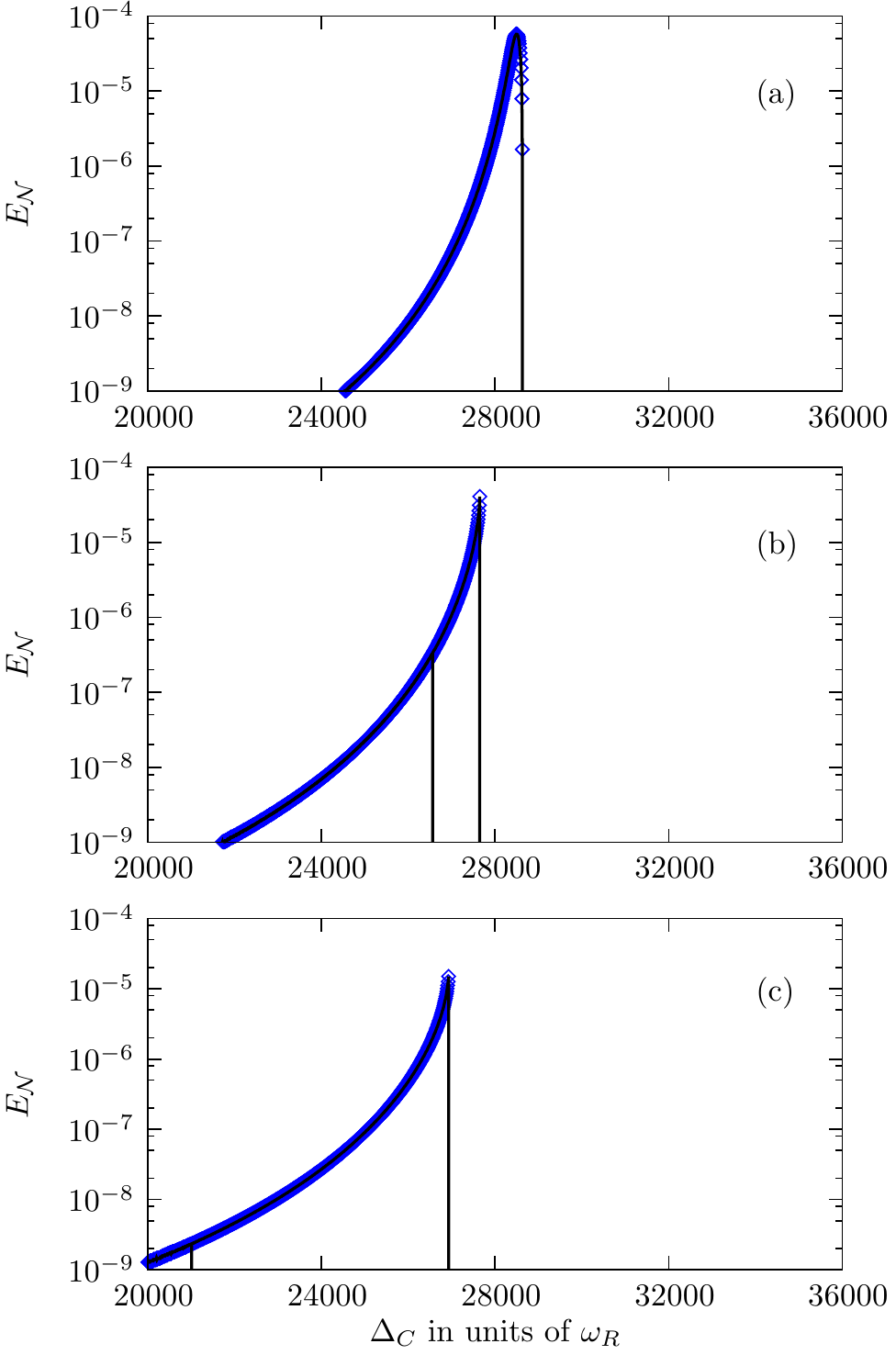}
  \caption{(color online) The logarithmic negativity of the condensate vs. the cavity detuning of the optomechanical model. All parameters are the same as for Fig. \ref{fig:phnum}.}
  \label{fig:logneg}
\end{center}
\end{figure}

To be able to relate our results more explicitly to other works, let us introduce quadrature operators, according to $\delta x=(\delta a+\delta a^\dagger)/\sqrt{2}$, $\delta y=-i(\delta a-\delta a^\dagger)/\sqrt{2}$, $\delta X=(\delta c+\delta c^\dagger)/\sqrt{2}$, $\delta Y=-i(\delta c-\delta c^\dagger)/\sqrt{2}$. These quadratude operators are hermitian operators, easily expressed with the help of the field operators $R$. We assemble them into the following row vector: $u=(\delta x, \delta y, \delta X, \delta Y)^T$. With the quadratures being hermitian, one can define a real correlation matrix by
\begin{equation} 
\label{eq:corrmatsymm}
C_{k,l}(t)=\frac{1}{2}\langle u_k(t)\,u_l(t)+u_l(t)\,u_k(t)\rangle,
\end{equation}
which is in the following block form
\begin{equation} 
\label{eq:cmblockform}
\mat{C}=\left[\begin{array}{c c}
\mat{P}&\mat{X}\\
\mat{X}^T&\mat{A}              
\end{array}\right],
\end{equation}
with $\mat{P}$ representing the correlations of the photonic degree of freedom, $\mat{A}$ representing atomic fluctuations, and $\mat{X}$ describing the cross correlations. For example, on top of the mean field contribution $|\alpha|^2$, there is a nonclassical part of the photon number which is given by
\begin{equation}
\label{eq:phnumnonclass}
n_{\text{ph}}'=\langle\delta a^\dagger\,\delta a\rangle=\frac{\langle\delta x^2\rangle+\langle\delta y^2\rangle-1}{2}=\frac{C_{1,1}+C_{2,2}-1}{2}.
\end{equation}
In an empty resonator close to zero temperature, where $\langle\xi^\dagger(t)\xi(t')\rangle=0$, the nonclassical part of the photon number \eqref{eq:phnumnonclass} vanishes, and the resonator field is in a pure coherent state. The fluctuation of the quadratures are distributed equally and the Heisenberg uncertainty principle is fulfilled in a sharp sense ($\langle\delta x^2\rangle=\langle\delta y^2\rangle=1/2$). However, due to atom-photon interaction, the photon fluctuations couple to those of the atomic motion. By iterative substitution of Eqs.\ \eqref{eqs:fluctom} one can simply check that $n_{\text{ph}}'=\langle\delta a^\dagger \delta a\rangle$ is no longer zero in the presence of a Bose-Einstein condensate. The photon field is no longer in a purely coherent state, and the correlation matrix $\mat{P}$ is not simply half of the unit matrix. Indeed, since $\delta a$ couples now to $\delta a^\dagger$ (through $\delta\tilde c$) and the coupling term is proportional with $\alpha_{\text{ss}}^2$, which is complex, the correlation matrix $\mat{P}$ is not isotropic. The angle of the major axis coincides with twice the phase of $\alpha_{\text{ss}}$, taking the values from $-\pi$ (when the system is far from the resonance) to zero (at resonance). By performing a rotation of the correlation matrix by this angle to bring it into a diagonal form, one of the eigenvalues will remain $1/2$, as in the pure coherent-state case, while the other one will always become bigger than $1/2$. Since the trace of a matrix is invariant under rotations, the nonclassical part of the photon number $n_{\text{ph}}'=(\lambda_>-1/2)/2$, with $\lambda_>$ being the bigger eigenvalue of $\mat{P}$. Figure \ref{fig:phnumnonclass} shows the nonclassical contribution to the photon number. The classical average $|\alpha_{\text{ss}}|$ of Fig. \ref{fig:phnum} is also plotted here with a dashed line for reference. It can be seen, by comparing the two curves, that the nonclassical contribution to the photon number is very small compared to the contribution of the mean field far from the resonance. When approaching the resonance,  the nonclassical part grows faster  than the mean field part,  and, in the vicinity of the resonance, it even exceeds the mean field contribution. The nonclassical contribution of the photon number is plotted only for the cavity cooling regime, where a steady state solution exists. 

Similarly, the number of atoms outside the condensate, i.e., the depletion, is evaluated as
\begin{equation} 
N'=\langle\delta c^\dagger\,\delta c\rangle=N\langle\delta\tilde c^\dagger\,\delta\tilde c\rangle=\frac{C_{3,3}+C_{4,4}-1}{2}.
\end{equation}
Figure \ref{fig:depletion} shows the steady-state number of particles outside the condensate for the parameter settings of the earlier plots. Notice that on the right hand side of the resonance depletion is not plotted. In this regime, where the cavity heats atomic motion instead of cooling it, there is no steady-state condensate, and depletion grows in time exponentially. The steady-state depletion is analogous to the quantum depletion of the ground state of a non-ideal Bose gas of atoms due to collisions.  In the cavity case the interaction between the atoms is provided by the collective coupling to the photon field. However, this depletion scales completely differently than that caused by s-wave scattering and is strongly influenced by the presence of the resonance.  The diffusion of atoms out of the condensate can also be interpreted as a quantum measurement-induced back action process which stems from the dispersive atom-light interaction \cite{murch08a,nagy09} and occurs also even in phase contrast imaging of a condensate where the photon field is propagating in free space \cite{leonhardt99,dalvit02b}. 
In a recent paper \cite{szirmai09a} we have shown that the depletion has a large steady-state value even in the limit of vanishing interaction strength $U_0$. For a fixed large detuning, $\Delta_C \gg \kappa,\, N U_0$, the amount of non-condensed atoms was estimated by $\Delta_C/\omega_R$, and is connected to the ratio of the photon energy and the energy of motional excitations. In Fig.\ \ref{fig:depletion} the detuning is a variable and the above condition is not fulfilled. However, far from resonance, the same ratio determines the depletion with the effective photon energy given by the detuning $\delta_C$.
%For a timescale much shorter than the characteristic time of the , the exponential functions in Eq.\ \eqref{eq:equtimecorr} can be expanded and one gets a linear growth of depletion in time.

The amount of entanglement between the atomic motional and photonic degrees of freedom can also be calculated with the help of the correlation function assuming that the state of the system is a Gaussian one. The logarithmic negativity, $\logneg$ is a useful measure of entanglement in our case, since it can be directly calculated with the help of the correlation matrix \eqref{eq:corrmatsymm}.
\begin{subequations} 
\begin{equation}
\label{eq:logneg}
\logneg=\mathrm{max}(0,-\ln2\eta^-),
\end{equation}
where
\begin{equation} 
\label{eq:numin}
\eta^-=2^{-1/2}\sqrt{\Sigma(\mat{C})-\sqrt{\Sigma(\mat{C})^2-4\det\mat{C}}}
\end{equation}
\end{subequations}
is the smaller symplectic eigenvalue of the two-mode Gaussian state, with $\Sigma(\mat{C})=\det \,\mat{P}+\det\mat{A}-2\det\mat{X}$. The state is an entangled state if and only if $\logneg\neq0$. The larger the logarithmic negativity, the larger the entanglement between the atomic motion and the photonic degree.

Figure \ref{fig:logneg} shows the steady-state value of the logarithmic negativity, $\logneg$ as a function of $\DC$ for the three parameter settings of Fig. \ref{fig:phnum}. These results hold only for the cooling solution. In the heating regime, where there is no steady-state, the entanglement between the photonic mode and the atomic motion also grows in time. The logarithmic negativity takes very small values in the whole range of the presented parameters except for a very small region around the instability. Apart from this narrow region, the entanglement is small even compared to the values of other optomechanical systems \cite{genes08a}. The smallness of the entanglement might be attributed to the big difference between the occupation numbers of the photonic and atomic modes, or equivalently, to the large difference in the effective energies of the decoupled subsystems. To reach higher values of entanglement either the timescales of the photonic and atomic degrees of freedom should be closer to each other, or a much stronger driving is needed to attain high photon numbers.

\section{Discussion}
\label{sec:disc}

In this paper we have investigated the one-dimensional dynamics of a Bose-Einstein condensate inside a driven optical cavity. As the dispersive atom-photon interaction couples the atomic motion to the dynamics of the photonic field in a nonlinear way, strong correlations can appear. % that are absent in the noninteracting system.
The strength of this coupling is inversely proportional to the detuning of the pump frequency from the atomic transition, therefore it can be tuned in experimental implementations. 
We recited and analyzed the mapping of the original system to a two mode effective model in which only the two highest populated one-particle states are kept from a plane-wave expansion of the atomic motion \cite{brennecke08a,murch08a,ritter09a}. By solving the coupled Gross-Pitaevskii equations it was possible to reproduce the bistable behavior caused by the nonlinear coupling \cite{brennecke07,ritter09a,zhang09a} and to provide a phase diagram of the system partitioning the whole parameter space into regions with full dynamical stability of the mean-field solution, the cooling region; to a region with a dynamically instability attributed to cavity heating; and to a region where both a stable and an unstable solutions can exist. We have compared the mean-field solution and the fluctuation spectrum of the optomechanical model to that of the model not restricted to the first two highly occupied modes. In the cavity heating region, the unstable polariton mode can have a positive imaginary part in the order of a kilohertz giving an evaporation rate of the Bose-Einstein condensate in milliseconds. Such timescale is in the experimental reach.

The dispersive atom-photon interaction not just causes cavity cooling or cavity heating but also alters the statistics of the constituent subsystems. In the framework of the optomechanical model, second order correlations were also investigated between the radiation field of the cavity and the motional mode of the Bose-Einstein condensate in the cavity cooling regime. Significant contributions beyond that of the mean field were found to the photon and particle numbers. The strong depletion of the condensate shows some analogy with the excess noise in lasers \cite{papoff08a,dalessandro09a}, already discussed in our previous work \cite{szirmai09a}. It is interesting however, that the huge nonclassical contribution in the autocorrelation of the photonic and atomic operators do not manifest in the entanglement of these variables. The lack of entanglement can be attributed to the big difference of the occupation of these modes. While the atom number was assumed to be in the order of $10^5$, the photon number ranged in the order of unity.  

%The issue on temperature. 1 excitation, estimation by thermal occupation\ldots

The experimental progress in combining cavity QED systems with ultracold atoms promises an interesting  playground to test the manifestation of light-matter interactions in the mesoscopic scale. In such systems both the radiation and the atomic part are dynamical entities. The better understanding of their interplay can have impact not just on our knowledge of nonequilibrium systems, but also on implementations of quantum information processing devices or quantum simulators of other systems.

\section{Acknowledgements}
\label{sec:ack}

The authors are grateful for Andr\'as Vukics, Claudiu Genes and Helmut Ritsch for useful discussions. We acknowledge funding from the National Scientific Fund of Hungary (NF68736, T077629), and from the National Office for Research and Technology (ERC\_HU\_09 OPTOMECH). G.~Sz. acknowledges Spanish MEC projects TOQATA (FIS2008-00784), QOIT (Consolider Ingenio 2010), ERC Advanced Grant QUAGATUA and EU STREP NAMEQUAM.

\end{document}